\documentclass[preprint,showpacs,preprintnumbers,amsmath,amssymb,prb,superscriptaddress]{revtex4}
\usepackage[dvips]{graphicx}
\usepackage{color}
\usepackage{bm}

\newcommand{\ket}[1]{\left| #1 \right\rangle}

\begin{document}

\title{Current-constraining variational approaches to quantum transport}

\author{P. Bokes } \email{Peter.Bokes@stuba.sk}
\affiliation{Department of Physics, Faculty of Electrical Engineering and
         Information Technology, Slovak University of Technology,
         Ilkovi\v{c}ova 3, 841 04 Bratislava, Slovak Republic}
\affiliation{Department of Physics, University of York, Heslington, York
          YO10 5DD, United Kingdom}
\author{H. Mera }
\affiliation{Department of Physics, University of York, Heslington, York
          YO10 5DD, United Kingdom}
\author{R. W. Godby}
\affiliation{Department of Physics, University of York, Heslington, York
          YO10 5DD, United Kingdom}
\date{\today{}}

\begin{abstract}
Presently, the main methods for describing a non-equilibrium
charge-transporting steady state are based on time-evolving it from the initial
zero-current situation.  An alternative class of theories would give
the statistical non-equilibrium density operator from
principles of statistical mechanics, in a spirit close to Gibbs
ensembles for equilibrium systems, leading to
a variational principle for the non-equilibrium steady state.
We discuss the existing attempts to achieve this using the maximum
entropy principle based on constraining the average current.
We show that the current-constrained theories result in a zero induced drop
in electrostatic potential, so that such ensembles cannot
correspond to the time-evolved density matrix, unless left- and right-going 
scattering states are mutually incoherent

\end{abstract}

\maketitle

\section{Introduction}
\label{sec-0}
The problem of charge transport through nanoscale objects
became a very intensive area of research during the last few years.
Numerical calculations reflect the rapidly advancing development
of experimental techniques exploring typical characteristics of atomic wires
or single-molecule junctions~\cite{Agrait03,Nitzan03}.  The most usual
theoretical methods used to address these issues are either direct
occupation of scattering states~\cite{Lang95,DiVentra00} or using the
non-equilibrium Green's function formalism for steady-state
(NEGF)~\cite{Taylor01,Brandbyge02}. Both
of these rely upon an assumption that the use of a ground-state
density-functional theory gives the effective self-consistent potential.
This assumption is now known not to be correct even though corrections
to it might be relatively small, which is particularly true for systems
with open channels~\cite{Sai04,Burke05}. 
The NEGF method makes use of semi-infinite reservoirs which demand
further approximations to the effective potential if one is to be able
to perform {\it ab initio} studies of the systems of interest,
e.g. typically the electronic structure of the leads are taken as that
of the leads in equilibrium.
The errors included this way can be assessed and calculations for
representative systems suggest them to be no more than a few
percent~\cite{Mera05}.
A second source of discrepancies
between the experimental and theoretical work comes from the ambiguity
of the geometry of the molecular junction one employs. The most common
choice is the ground state geometry. Even though several interesting
results concerning non-equilibrium forces exist~\cite{DiVentra04,DiVentra02},
{\it ab initio} studies of current-distorted geometries are still absent.

Almost all of the above-mentioned development is based on
time-evolution as a means to obtain the non-equilibrium steady
state~\cite{Gebauer04b,Stefanucci04,Kurth05}, although in most cases
the time evolution is used only formally for the derivation of the formulae
used within the NEGF formalism.
The other alternative would be to use some sort of variational principle
that would directly lead us to the non-equilibrium density matrix (or state).
While it might be difficult to believe that such a principle would exist
for a general dissipative system, its existence for the purposes of quantum
transport is relatively easy to accept. At the level of a non-interacting
or mean-field description of the electrons,
it has been known for some time that such a variational principle
exists~\cite{Ng92,Todorov00}, and it can be extended formally to fully
interacting electron systems, provided that there is a physical relaxation
process that ``washes out'' initial correlations~\cite{Hershfield93}.
These are based on the search for a state with minimum energy, consistent
with prescribed numbers of right- and of left-going electrons
(hereafter referred to as the ``left-right'' scheme, LRS) or prescribed 
average current (current constrained approaches). 
The LRS prescription is directly motivated by the Landauer-B\"{u}ttiker formalism~\cite{Buttiker85}
developed for mesoscopic systems and can be generalized by maximizing the information
entropy with constraints on the average energy (corresponding to
the introduction of the temperature in equilibrium) and average numbers of left- and right-going particles
To remove the concept of single-particle orbitals from this approach,
Frensley suggested to constrain the local Wigner distribution functions
for electrons with positive and negative momenta in the left and right lead
respectively~\cite{Frensley90}. This has the advantage that it can be used
also for interacting approaches as has been demonstrated recently
by Delaney and Greer~\cite{Delaney04}. A different extension of the
single-particle theory was presented by
Hershfield~\cite{Hershfield93}, who formally constructed a non-equilibrium
steady-state density matrix using many-body field operators corresponding
to generalizations of the scattering states.

As a alternative to the density constraint, several authors have suggested the
total current as a means to keep the steady state out of
equilibrium~\cite{Ng92, Heinonen93,Johnson95,Kosov01,Kosov02,Kosov03,Bokes03,
DiVentra04b}. The current
constraint, unlike the density constraint, has immediately a well-defined form
even for interacting electrons. It was
soon realized that the current-constrained density matrix corresponds
to a situation different to that obtained within the LRS,
although no clear consensus exists regarding what experimental situation it
describes.
It has been mentioned already in the work of Ng~\cite{Ng92}
that the theory should correspond to a constant-current experiment (as opposed
to constant-voltage). However, this cannot be the only criterion,
since the system of interest is very small and so the ability of the system
to explore the whole Hilbert space of admissible density matrices should
be considered with care. The latter property is indeed at the center
of the formulation employing constrained searches within the maximum entropy
principle~\cite{Bokes03,Jaynes78}. In this respect it is interesting
to note recent work by Di Ventra and Todorov~\cite{DiVentra04b}
who have formally considered a variational principle based on constraining
the current for a quasi-steady state of discharge of a large but finite
electrodes through a nanojunction. The steady-state-current constrained
ensembles, which are the subject of the present paper, should correspond to a
long-time and infinite-size limit of their considerations.

The various current-constraint formulations differ in some details.
Ng~\cite{Ng92} considers a treatment where the current operator is altered in
such a way that it does not mix the right- and left- going scattering states.
To achieve this one has to drop all the off-diagonal matrix elements
of the current operator in the scattering-states representation.
This eventually leads to a theory that is similar to the LRS,
but with the occupancies now depending on
the current that the particular state carries, as well as its energy. The
applied bias is determined
from the difference in the local electrochemical potential between
the left and right asymptotic regions which, strictly speaking, corresponds
to the electrostatic drop around the sample.
Heinonen and Johnson~\cite{Heinonen93,Johnson95} consider only systems
that are translationally invariant along the current flow. Under such
circumstances the off-diagonal matrix elements of the current operator
are absent by symmetry and the notion of applied bias is only formal.
They determine the latter from the analogy with the
scattering-states-occupation
theory and the resulting $I-V$ characteristics in the linear
regime are identical to the LRS description. One should point out
that this analogy fails as soon as one enters a strongly non-linear $I-V$
regime with, for example, a current flow through a resonant barrier.

In both of these treatments the effective Hamiltonian
with the current constraint commutes with the physical Hamiltonian or, in
other words, that the constrained density matrix is stationary. The
approach developed by Kosov~\cite{Kosov01,Kosov02} departs from this point
and instead constrains the current to be uniform throughout the system.
As a result, the density matrix
is not time-independent which brings into question
its relevance for the description of a steady state.
The problem of the steady-state character of the density matrix obtained from
a current-constrained search has been studied by the authors of this
paper~\cite{Bokes03}. It has been found that the steady-state requirement
{\it does} remove most of the off-diagonal matrix elements of the current
operator, as is assumed by Ng.  However, those off-diagonal elements between
states with the same energy do {\it not} disappear, which leads to a
density matrix different from that anticipated by Ng. The induced drop in the
potential
was found from the local neutrality
conditions in the asymptotic regions, in a manner similar to that of Ng, and
the paper additionally discussed the
relation of the Lagrange multiplier $A$ (that imposes the current constraint)
to the applied bias voltage.

The aim of this paper is to present a clear relationship between the
above-mentioned current-constraining schemes, as well as to discuss
their limitations.  We start by analyzing the role of the external applied
bias within the formalism. This yields a link between the time-evolved and
variational approaches, and gives a supportive argument for
the form of the steady-state requirements implemented in our earlier
work~\cite{Bokes03}. In  Section \ref{sec-2} we address the induced drop
in the electrostatic potential. Conversely to what has been claimed before
we have found that
unless we remove {\it all} of the off-diagonal elements of the current
operator, the induced drop is exactly zero. This holds for the uniform-current
approach, discussed in the Section \ref{sec-3}, where all elements
are retained~\cite{Kosov01} as well as for the approach (used in our previous
study~\footnote{In our original treatment we have found a nonzero drop due
to an error in the derivation of the induced density, i.e. Eq.~(14)
in~\cite{Bokes03} is incorrect.}) in which the current operator has
off-diagonals only between equal-energy states. In Section \ref{sec-2b}
we describe how many-body interactions of the electrons with the environment
correct the previous results to a physically meaningful picture where
nonzero drop in the potential is found and the agreement with the
conventional approaches (NEGF, LRS) achieved within the linear response regime.
\section{Applied External Field and Maximum Entropy}
\label{sec-1}

We describe the steady-state-current situation using the maximum
entropy principle for the statistical density operator
$\hat{\rho}$~\cite{Jaynes78}
\begin{equation}
         \delta \left\{ - \mathrm{Tr} \left[ \hat{\rho} \log( \hat{\rho} )
                         \right] - \sum_i \mathrm{Tr} \lambda_i
                         \left[ \hat{\rho} \hat{A}_i \right]
                 \right\} = 0,
\end{equation}
where $\lambda_i$ are Lagrange multipliers that guarantee chosen values, $A_i$,
of the averages
of chosen operators $\hat{A}_i$.
That is, instead of following the specific time evolution caused by an applied
external field, we assume that the final steady-state is representable
by the statistical density operator $\hat{\rho}$ that possess the same
current and/or induced drop in potential. While all of the existing schemes
agree on using the total energy and total number of electrons as two of the
constrained operators $\hat{A}_i$, the constraint that keeps the system out of
equilibrium
varies and is either the total current or the occupancies of right- and
left-going scattering states.

In real-time evolution approaches, there are no non-equilibrium constraints or
multipliers.
Instead, the applied external field acts as a driving force for the current
flow. The character of the external field seems to be a source of certain
confusion in the community. Some authors use a ramp-like external potential
that has a finite slope between contacts~\cite{Kosov02,Delaney04} even within
maximum entropy schemes, while some avoid its presence
altogether~\cite{Ng92, Heinonen93, Johnson95, Hershfield93, Bokes03}.
On the other hand, in calculations based on
the time evolution (scattering states, NEGF formalism), the role
of the applied external field is frequently circumvented by the application of
a difference in
electrochemical potential between two initially isolated leads.  The electrons
are taken to be non-interacting while in the leads.
This construction, however, leads to violation of
local charge neutrality in the non-interacting regions as discussed
elsewhere~\cite{Mera05}.  Neither does it clarify
the relation between the total and induced electric field. Clearly,
the problem with the applied external field can be tracked down to the fact
that one tries to model the effect of the battery within the calculation.

The appropriate form of the applied field is available from considerations
originally made in the linear response regime~\cite{Bokes04}. To have a system
infinite along
the direction of flow and characterized by a finite drop in external potential,
$\Delta V$, one needs to consider the large-time limit of a field
\begin{equation} \label{eq-ext-field}
         E^{ext}(x,t) = \frac{\Delta V}{2 u t}( \theta(x+ut) - \theta(x-ut) ),
\end{equation}
where $\theta(x)$ is the unit step function, $x$ is the direction of current
flow and $u$ is the desired speed of the front between the region with
an applied field and that
without field. This represents a situation that has a constant drop $\Delta V$
in the potential at all times $t$ and, for sufficiently large $u\gg v_F$,
it produces a uniform current even in the case of interacting electrons,
i.e. the field is not screened out. The steady state is obtained in
the $t\rightarrow \infty$ limit where the drop $\Delta V$ persists, but
the external field in any finite part of system is zero.
(e.g. $ E(x=0,t) = \Delta V / (ut) \rightarrow 0$). The initial localization
of the field-containing region leaves no long-time signature other than
the steady current that flows, and associated changes in the electronic
structure such as density or induced potential.

The above considerations of the steady state show that the Hamiltonian
{\it with zero applied field} must be used when constructing the density
operator within the maximum entropy ansatz.  Similarly, we believe that
the use of a finite external field together with the maximum
entropy prescription is simply incorrect, and its application in other
calculations should be reconsidered. The {\it induced} field {\it will},
however, appear in the calculation as a consequence of the current constraint
applied to the density matrix.

The second outcome of this observation concerns the steady-state
character of the system. Once we accept that the Hamiltonian
present is that without the applied field, the stationarity of the
statistical density operator requires~\cite{Bokes03}
\begin{equation}
         \left[ \hat{\rho} , \hat{H} \right] = 0 \label{eq-steady-state-cond}
\end{equation}
This identity has to be included when performing the constrained search
for the operator $\hat{\rho}$. At the same time, it should be clear
that this condition can be fulfilled only for a system infinite along
the direction of the current flow. (The only exception is a system of
perfect translational invariance and finite periodic boundary conditions.
As this represents a very special and highly non-generic case
-- an arbitrarily small perturbing potential spoils the perfect translational
invariance and therefore the ability of the system to carry current --
we will not be concerned with it in our further discussion.)

The condition (\ref{eq-steady-state-cond}) has been shown to follow also
from a time-evolution point of view by Hershfield~\cite{Hershfield93}.
In this work it is also correctly pointed out that, as opposed to
the equilibrium expectation value, non-equilibrium systems
are characterized by an effective Hamiltonian that enters the statistical
density operator, which is different from the true physical Hamiltonian
characterizing the time evolution or time-correlations in the system.
This, as we will see, significantly complicates the formal
development of the theory for interacting non-equilibrium steady-state systems.

\section{The invariant current approach and the induced drop in the potential}
\label{sec-2}

In the invariant-current approach the constraint that keeps the system
out of equilibrium is chosen to be the current at a particular point, $x_0$
\begin{equation}
         I(x_0) = \int d \vec{S} \cdot \mathrm{Tr} \left[
                 \hat{\rho} \hat{j}(x_0,y,z) \right],
\end{equation}
where $\hat{j}(x,y,z)$ is the operator of the current density at
${\bf r}=(x,y,z)$.
As we have shown~\cite{Bokes03}, this requirement, together with the
steady-state restriction~(\ref{eq-steady-state-cond}), leads to the statistical
operator
\begin{equation}
         \hat{\rho} = \exp\{\Omega - \beta ( \hat{H} - \mu \hat{N}
         - A \hat{I}^{0} ) \}. \label{final-DM}
\end{equation}
where the operator $\hat{I}^{0}$ is the invariant part of
the current operator~\cite{Kubo59,Grandy88}
\begin{equation}
         \hat{I}^{0} = \lim_{T \rightarrow \infty } \frac{1}{2T} \int_{-T}^{T}
         \hat{I}(t) dt,
\end{equation}
which is independent of the position of current measurement $x_0$.
This arises from the fact that for a stationary density matrix the current
fulfills the continuity equation $\nabla \cdot \hat{\bf j} = - \dot{n} = 0$.
The time-dependence of the operator $\hat{I}(t)$ is determined by
the Hamiltonian of the system $\hat{H}$ which, similarly to
the case of scattering field operators in Hershfield's
work~\cite{Hershfield93},
hinders the development of the theory for interacting electrons.

At the mean-field level of approximation, the one-particle density matrix
can be used to obtain all required quantities.
Instead of the many-body Hamiltonian we need to consider the
effective one-particle Hamiltonian given by~\footnote{We treat
only the 1D model for simplicity; the generalization of the presented
results for several subbands is straightforward.}
\begin{equation}
         \hat{h} = -\frac{1}{2} \partial^{2}_{x} + V(x) - \mu -
                 A I^{0}(x), \label{eq-h1}
\end{equation}
where $V(x)$ is the self-consistent potential, $\mu$ chemical potential,
$A$ a Lagrange multiplier belonging to the current constraint
and $I^0$ the invariant current operator. Even though we eventually obtain
the one-particle density matrix that closely resembles the usual local
Fermi distribution, its derivation is non-trivial. The problem arises from
the fact that our system cannot, strictly speaking, be obtained as a limit
of a finite one. First, the finite drop in the potential, $\Delta \phi$,
makes it impossible to consider periodic boundary conditions, and, second,
the existence of a non-zero current flow hinders the construction of hard walls
placed at finite, but large distances at right and left, as used by
Adawi~\cite{Adawi64}. Similarly, the use of periodic boundary conditions, as
implicitly appear in some current-constraint-based
treatments~\cite{Ng92,Heinonen93,Johnson95}, is not consistent with the
possible existence of an overall drop in the electrostatic potential and
the non-zero current. For these reasons we give its detailed derivation
in the Appendix \ref{sec-app-1}. The resulting one-particle density matrix is
\begin{equation}
         n(x,x') = \label{eff-Fermi}
         \sum_{\alpha} \int dE \frac{\chi_{E,\alpha}(x)
         \chi^{*}_{E,\alpha}(x')}{e^{\beta \left( \tilde{E}_{\alpha}(E)
                 - \mu \right) } + 1},
\end{equation}
where $\chi_{E,\alpha}(x)$ are energy-normalized states that diagonalize
the Hamiltonian~(\ref{eq-h1}) and $\tilde{E}_{\alpha}(E)$ are the corresponding
eigenvalues. The latter can be expressed in the basis of scattering states
of the physical Hamiltonian (i.e. without the term containing the current
operator
in Eq.~(\ref{eq-h1}))
\begin{eqnarray}
         \chi_{E,\alpha}(x) &=& \sum_{\nu}
                 \phi_{E,\nu}(x) U_{\nu,\alpha}(E),
\end{eqnarray}
where $\phi_{E,\nu}(x), \nu=R,L$ represent right- or left-going scattering
states at the energy $E$ given asymptotically as
\begin{eqnarray}
        \phi_{E,R}(x>>0) &=&
                 \frac{1}{\sqrt{2 \pi k } }
                 t \mathrm{e}^{iq x}, \label{eq-scatt-r} \\
        \phi_{E,L}(x<<0) &=&
                 \frac{1}{\sqrt{2 \pi k } }
                 \tilde{t} \mathrm{e}^{-ik x}, \label{eq-scatt-l}
\end{eqnarray}
where $k=\sqrt{2E}$ and $q=\sqrt{2(E+\Delta \phi)}$ are the wave-vectors
on the far left and far right respectively and $t$ and $\tilde{t}$ are
transmission
amplitudes for right-going and left-going electrons respectively.
($t$ and $\tilde{t}$ depend on the energy but we will not write this dependence
explicitly.)
We note that the states $\chi_{E,\alpha}$ can be labeled with the energy
because the effective Hamiltonian commutes with the physical one, as required
by the stationarity condition~(\ref{eq-steady-state-cond}). From this
and a glance at Eq.~(\ref{eq-h1}) it follows that the states $\chi_{E,\alpha}$
are simultaneously eigenstates of the invariant current operator $I^0$.
Therefore the second index differentiates between energy-degenerate states
which, in the simplest case of a single 1D channel considered here, attains
two different values, '+' and '-', depending on the sign of the invariant
current eigenvalue of the respective state. In Appendix \ref{sec-app-2} we
show that the states $\chi_{E,+}$ and $\chi_{E,-}$ transform one into another
under the time-reversal $T$,
\begin{equation}
         T \chi_{E,+} = \chi^*_{E,+} = e^{i\phi} \chi_{E,-}.
\label{eq-time-rev}
\end{equation}
The latter relation has an important consequence for the induced
change in the density in the linear regime, as we will discuss below.

We have already indicated that the scattering-states representation plays
a fundamental role not only in the LRS approach but also in the
current-constraint schemes. It is therefore useful to express the
current operator in the scattering states representation~\cite{Buttiker92}
\begin{eqnarray}
         2\pi\frac{dI_{\nu,\eta}(x<<0)}{dE} &=&
         \delta_{\nu,1} \delta_{\eta,1} - S^{\dagger}_{\nu,1} S_{1,\eta}
         = \left[ \begin{array}{cc}
                 1 - |r|^2 & - r^* \tilde{t} \\
                - r \tilde{t}^* & - \frac{k}{q} |\tilde{t}|^2
                 \end{array} \right], \label{eq-curr-1} \\
          2\pi\frac{dI_{\nu,\eta}(x>>0)}{dE} &=&
         - \delta_{\nu,2} \delta_{\eta,2}+S^{\dagger}_{\nu,2} S_{2,\eta}
         = 2\pi\frac{dI_{\nu,\eta}(x<<0)}{dE}, \label{eq-curr-2}
\end{eqnarray}
where $S_{\nu,\eta}$ is the scattering matrix
\begin{equation}
         S_{\nu,\eta} = \left[ \begin{array}{cc} r & \tilde{t}  \\
                 t & \tilde{r} \end{array} \right]
\end{equation}
and the last equality in Eq.~(\ref{eq-curr-2}) follows from the unitarity
of $S_{\nu,\eta}$ (current conservation) and $t, r$ and $\tilde{t}, \tilde{r}$
are the transmission and reflection amplitudes for right- and left- going
electrons. Similarly to the latter, we do not write explicitly the energy
dependence of the $S$-matrix or the current operator matrix. Since
the invariant current operator $I^0$ in (\ref{eq-h1}) is related to the matrix
of the current operator multiplied by a delta-function of
energy~(see Appendix \ref{sec-app-1}),
\begin{equation}
         I^0(E,\alpha;E',\beta) \times A
         = \frac{d I_{\alpha,\beta}}{dE} \delta(E-E') \times \tilde{A},
\end{equation}
the states $\chi_{E,\alpha}(x)$ automatically diagonalize the kinetic
and potential energy terms in Eq.~(\ref{eq-h1}). To diagonalize
the complete effective Hamiltonian they need to diagonalize also the
current operator which is given by Eq.~\ref{eq-curr-1}.
The unitary transform that achieves this has been found before~\cite{Bokes03}
\begin{equation}
         U_{\nu \alpha} = \left[  \begin{array}{cc}
         \frac{r^* \tilde{t}/|t| }{\sqrt{2(1-|t|)}} &
         \frac{r^* \tilde{t}/|t| }{\sqrt{2(1+|t|)}} \\
         \frac{|t|-1}{\sqrt{2(1-|t|)}} &
         \frac{|t|+1}{\sqrt{2(1+|t|)}}
         \end{array} \right] \label{eq-U}
\end{equation}
with the corresponding effective eigenvalues
\begin{equation}
         \tilde{E}_{\alpha}(E) = E \pm \tilde{A} |t|  \label{eq-eff-E}.
\end{equation}
The calculation of the current is then straightforward using the one-particle
density matrix,
\begin{equation}
         \langle I(x) \rangle = \sum_{\alpha,\nu,\eta} \int dE
         \frac{1}{e^{\beta \left( \tilde{E}_{\alpha}(E) - \mu \right) } + 1}
         U^\dagger_{\alpha,\nu} \frac{dI_{\nu,\eta}(x)}{dE}
         U_{\eta,\alpha} \label{eq-av-current}
\end{equation}
Using this expression we find the dependence of the current on the
renormalized Lagrange multiplier $\tilde{A}$.

For comparison with other approaches as well
as experiments, we also need the dependence of the current on the drop in the
electrostatic potential, $\Delta \phi$, which can be obtained from the induced
change in the density $\delta n(x)$ via the expression
\begin{eqnarray}
         \phi(x) &=& \int v(x,x') \delta n(x') dx', \nonumber \\
         \Delta \phi &=& \phi(+\infty) - \phi(-\infty), \label{eq-Poisson}
\end{eqnarray}
where $v(x,x')$ is an appropriate effective electron-electron interaction.
Alternatively, $\Delta \phi$ can be obtained from the local
neutrality conditions in the right and left electrodes~\cite{Mera05} for
the self-consistently determined scattering matrix (hence the link to
the resistivity dipoles based formula). The local neutrality conditions demand
\begin{eqnarray}
         -n(x<<0) + n_{B} &=& 0 \\
         -n(x>>0) + n_{B} &=& 0,
\end{eqnarray}
where $n_{B}$ is the positive charge density of the background, assumed
to be the same in both electrodes for simplicity. Subtracting these two we
get a single condition
\begin{equation}
         \Delta n = n(x<<0) - n(x>>0) = 0  \label{eq-neutrality}
\end{equation}
This identity alone can be used to determine the drop in the potential
$\Delta \phi$, for given values of $\tilde{A}$  and $\mu$. 
The density in the asymptotic regions is obtained in a way
similar to the total current: we first find the expressions for the matrix
of the density in the asymptotic regions within the scattering states
representation
\begin{eqnarray}
         \frac{dn_{\nu \eta}(x<<0)}{dE} &=&
         \frac{1}{2\pi k} \delta_{\nu,1} \delta_{\eta,1} + \frac{1}{2 \pi k}
         S^{\dagger}_{\nu,1} S_{1,\eta}
         = \frac{1}{\pi k} \delta_{\nu,1} \delta_{\eta,1} - \frac{1}{k}
         \frac{d I_{\nu \eta}(x<<0)}{dE} \label{eq-dn-ll} \\
         \frac{dn_{\nu \eta}(x>>0)}{dE} &=&
         \frac{1}{2\pi q} \delta_{\nu,2} \delta_{\eta,2} + \frac{1}{2 \pi q}
         S^{\dagger}_{\nu,2} S_{2,\eta}
         = \frac{1}{\pi q} \delta_{\nu,2} \delta_{\eta,2} + \frac{1}{ q}
         \frac{d I_{\nu \eta}(x>>0)}{dE}. \label{eq-dn-rr}
\end{eqnarray}
Second, we express the local neutrality condition (\ref{eq-neutrality})
using the density matrix
\begin{equation}
         \sum_{\alpha,\nu,\eta} \int dE
         \frac{1}{e^{\beta \left( \tilde{E}_{\alpha}(E) - \mu \right) } + 1}
         U^\dagger_{\alpha,\nu} \left[
         \frac{dn_{\nu,\eta}(x<<0)}{dE}-\frac{dn_{\nu,\eta}(x>>0)}{dE} \right]
         U_{\eta,\alpha} = 0 . \label{eq-neutrality-dn}
\end{equation}
To obtain algebraic results we need to restrict our treatment to
the $\beta \rightarrow \infty$ limit (corresponding to the density matrix
with the minimal energy for given constraints). Under these circumstances
the effective Fermi distribution takes the form
\begin{equation}
         \frac{1}{e^{\beta \left( \tilde{E}_{\alpha}(E) - \mu \right) } + 1}
         \approx \delta_{\alpha,1} \theta(\mu - E + \tilde{A} |t(E)| )
                 + \delta_{\alpha,2} \theta(\mu - E - \tilde{A}|t(E)|),
                 \label{eq-step}
\end{equation}
where the two step functions $\theta()$ correspond to the positive and negative
eigenvalues in Eq.~\ref{eq-eff-E}. Assuming $\tilde{A}>0$, we see that starting
from $E>\mu_2=\mu-\tilde{A}|t(\mu_2)|$ and up to
$E<\mu_1=\mu+\tilde{A}|t(\mu_1)|$,
only one of the two degenerate states $\chi_{E,\alpha}$ will be
occupied~\footnote{This is strictly valid only if $\tilde{A}$ is smaller than
the range of energies where $|t(E)|$ varies significantly}.

For $E<\mu_2$ the contributions to the density clearly can not depend
on the unitary rotation between the scattering states and we therefore
have
\begin{eqnarray}
         \sum_{\alpha} \frac{dn_{\alpha,\alpha}}{dE}(x<<0) &=&
         \sum_{\alpha,\nu,\eta}
         U^\dagger_{\alpha,\nu} \frac{dn_{\nu,\eta}}{dE}(x<<0) U_{\eta,\alpha}
         = \frac{1}{\pi k}, \label{eq-dn-r} \\
         \sum_{\alpha} \frac{dn_{\alpha,\alpha}}{dE}(x>>0) &=&
                 \frac{1}{\pi q}. \label{eq-dn-l}
\end{eqnarray}
The contribution from the singly occupied state $\chi_{E,\alpha=1}(x)$
we find from Eqs.~(\ref{eq-dn-ll},\ref{eq-dn-rr},\ref{eq-U}) with
a surprisingly simple result
\begin{eqnarray}
         \frac{dn_{11}}{dE}(x<<0) &=& \sum_{\nu,\eta}
         U^\dagger_{1,\nu} \frac{dn_{\nu,\eta}}{dE}(x<<0) U_{\eta,1}
                 = \frac{1}{2 \pi k}, \label{eq-dn-1l} \\
         \frac{dn_{11}}{dE}(x>>0) &=& \frac{1}{2 \pi q}. \label{eq-dn-1r}
\end{eqnarray}
Finally combining Eqs.~(\ref{eq-step},\ref{eq-dn-l},\ref{eq-dn-r},
\ref{eq-dn-1l},  \ref{eq-dn-1r}) in the Eq.~(\ref{eq-neutrality-dn})
we get
\begin{equation}
         \int_{0}^{\mu_2} \frac{dE}{\pi k} + \int_{\mu_2}^{\mu_1}
         \frac{dE}{2\pi k}
         - \int_{-\Delta \phi}^{\mu_2} \frac{dE}{\pi q} -
         \int_{\mu_2}^{\mu_1} \frac{dE}{2 \pi q} = 0.
\end{equation}
The last equation can be satisfied only for $\Delta \phi =0$.
This means that the invariant-current scheme with the off-diagonal
elements at the same energy retained leads to {\it no induced potential drop}
and therefore its applicability to common nano-contacts is doubtful, even in
a constant-current experiment.

Qualitatively, this result is a consequence of the fact that the
contribution to the density per-energy of each doubly degenerate
state $\chi_{E,\alpha}$, $dn_{\alpha,\alpha}/dE$, is almost the same~\footnote{
It is straightforward to show that the small difference in the contribution
to the densities $\sim \frac{1}{q} - \frac{1}{k}$ is not sufficient
to remedy the problem of local charge neutrality that appears for $\Delta \phi
\neq 0$.}
far right and far left (see Eqs.~(\ref{eq-dn-1l},\ref{eq-dn-1r})).
If we assume that the right-electrode, having the same background charge
density
as the left one, has the bottom of its local density of states at
$E=-\Delta \phi$, below that of the left electrode ($E=0$), occupying
this portion of the energy spectrum ($E<0$) will partially compensate
for the background charge in the right electrode. Once the states below
the bottom of the left electrode are filled, adding each electron into
next state $\chi_{E,\alpha}$ will contribute in both electrodes almost
equally so that when attempting to compensate for the background charge
of the left electrode we will inevitably overload the right electrode or,
when neutralizing the right electrode there will not yet be enough electronic
charge in the left one. (We would like to stress that the above
reached conclusions are valid for $\beta \rightarrow \infty$ limit, finite
$\tilde{A}$ and a regime, in which the equations for $\mu_{1}$ and $\mu_{2}$
have a unique solution. The latter fails to be fulfilled if either $\mu_{1}$ or
$\mu_{2}$ approaches a resonant energy level of the potential $V(x)$.)

An even more surprising result appears in the linear-response regime,
i.e. when $A I^0 << \mu$. Under these circumstances
the induced change in the density $\delta n(x)$,
$$
\delta n(x) = n(x,x)|_{\tilde{A}} - n(x,x)|_{\tilde{A}=0},
$$
where $n(x,x)|_{\tilde{A}}$ is the diagonal of the density matrix given in the
Eq.~(\ref{eff-Fermi}) for a given (small) value of the renormalized Lagrange
multiplier $\tilde{A}$, is simply given by
\begin{equation}
         \delta n(x) = 2 \tilde{A} |t(\mu)| \left( |\chi_{+}(x)|^2 -
                 |\chi_{-}(x)|^2 \right) \label{eq-delta-n} +
                 \mathcal{O}(|\tilde{A}|^2).
\end{equation}
Using the time-reversal character of the states $\chi_{E,\alpha}$,
Eq.~(\ref{eq-time-rev}), we immediately obtain that
$\delta n(x) = \mathcal{O}(|\tilde{A}|^2)$, i.e. there is no change
of the density in the linear regime. Using this result in the
formula for the induced drop, Eq.~(\ref{eq-Poisson}), we once again
confirm the above obtained result (valid even for finite $\tilde{A}$) of zero
induced drop in electro-static potential.

\section{The invariant current approach with decoherence}
\label{sec-2b}

The situation is quite different when the off-diagonals of the
invariant current operator are dropped. This  makes the density
matrix (\ref{eff-Fermi}) identical to that given by Ng~\cite{Ng92}.
The elimination of the off-diagonals can be made physically plausible
from the assumption of phase independence between the right- and left-
going scattering states. Namely if we ascribe an independent fluctuating phase
$e^{i\theta_{\nu} }$ to each of these two, the off-diagonal matrix elements
of the one-particle density matrix, Eq.~(\ref{eff-Fermi}), are easily seen
to be averaged to zero. The effect of the coupling to the environment
is to suppress, through averaging over a fluctuating phase, any mixing of
left- and right- going states in the pure states that are summed
in the density matrix. In effect, these pure states are constrained to
be either left- or right- going. This means that the off-diagonal elements
of the current operator no longer play any role in the determination of
the steady-state density matrix.

To recast the above considerations into a more formal language we need
to include a description of the effect of the environment in the 
Green's-function-based derivation of the one-particle
density matrix given in Appendix (\ref{sec-app-1}).
The Green's function, that directly leads to the one-particle density matrix,
is defined in terms of the time-ordered product of
the field operators,
\begin{eqnarray}
         G(2,1) = - \mathrm{Tr} \left[  \hat{\rho} T\{ \psi(2) \psi^\dagger(1)
\}
                          \right],
\end{eqnarray}
where $\hat{\rho}$ is the many-body density matrix that specifies
occupations of the states of the whole system, i.e. the degrees of freedom
of the environment that cause the above discussed fluctuating phases.
Rewriting the Green's function in terms of creation and annihilation operators
of electrons in the scattering states,
\begin{eqnarray}
         G(2,1) = - \int dE dE' \sum_{\nu,\eta}
                 \phi_{E,\nu}(x) \phi^*_{E',\eta}(x') \mathrm{Tr}
                 \left[ \hat{\rho} T\{ c_{E,\nu}(t_2)
c^\dagger_{E',\eta}(t_1)\}
                 \right],
\end{eqnarray}
we observe that the Green's function will have {\it only} the diagonal elements
with respect to the $\nu,\eta=R/L$ index, i.e.
\begin{equation}
         \mathrm{Tr}
                \left[ \hat{\rho} T\{ c_{E,\nu}(t_2) c^\dagger_{E',\eta}(t_1)\}
                \right] \approx \mathrm{Tr}\left[
                 \int \frac{d \theta_\nu d \theta_\eta}{4\pi^2}
                \hat{\rho}_e e^{i\theta_\nu} e^{-i\theta_\eta}
                 T\{ c_{E,\nu}(t_2) c^\dagger_{E',\eta}(t_1)\} \right]
                 \sim \delta_{\nu, \eta},
\end{equation}
where $\hat{\rho}_e$ is a many-electron density operator already {\it without}
the environment's degrees of freedom which were effectively
taken into account via averaging over the phases of the right- and left-
going scattering states.
From this it follows that the equation of motion for this Green's function,
Eq.~(\ref{eq-of-motion-RL}), is already in a diagonal form and so is the
one-particle density matrix $n(x,x')$, Eq.~(\ref{eff-Fermi-app}).

Anticipating the sources of decoherence and therefore the irrelevance
of the invariant current operator's off-diagonal elements, we can proceed
rather straightforwardly in a derivation of Ng's~\cite{Ng92} as well as our
former results~\cite{Bokes03} for the theory with decoupled right- and
left- going states.  Since the current matrix~(\ref{eq-curr-1}) is effectively
in a diagonal form, we can obtain the results considering the transform
$U_{\nu,\alpha} = \delta_{\nu,\alpha}$, so that the states
$\chi_{E,\alpha}(x)$, entering the one-body density matrix~(\ref{eff-Fermi}),
are directly the scattering states. The effective eigenvalues prescribing
the occupations are then given by
\begin{equation}
         \tilde{E}_{\alpha} (E) = E \pm \tilde{A} |t(E) |^2.
\end{equation}
What comes as an essential difference, as compared to the case when
the decoherence is not accounted for, is that the contribution to the
density from the singly-occupied states now gives
\begin{eqnarray}
         \frac{dn_{11}}{dE}(x<<0) &=& = \frac{2-|t(E)|^2}{2 \pi k},
\label{eq-dn-1ln} \\
         \frac{dn_{11}}{dE}(x>>0) &=& \frac{|t(E)|^2}{2 \pi q},
\label{eq-dn-1rn}
\end{eqnarray}
as can be easily seen by inspection of the expression~(\ref{eq-dn-ll}),
(\ref{eq-dn-rr}), (\ref{eq-curr-1}). Clearly, the contributions to
the local densities far right and far left are now significantly different
and therefore the qualitative argument for zero induced drop in the potential
does not apply.
Using expressions (\ref{eq-dn-1ln}),(\ref{eq-dn-1rn}) in the local neutrality
condition~(\ref{eq-neutrality}) eventually leads to the linear response result
\begin{equation}
         \Delta \phi = 2 \tilde{A} |t(\mu)|^2 |r(\mu)|^2 .
\end{equation}
Using equations (\ref{eq-av-current}), (\ref{eq-step}) and
(\ref{eq-curr-1}) we obtain the current in the linear response
\begin{equation}
         \langle I \rangle = 2 \tilde{A} |t(\mu)|^2 \frac{|t(\mu)|^2}{2\pi}.
\end{equation}
and therefore combining the last two expressions we arrive at the
well known result~\cite{Buttiker85}
\begin{equation}
         \frac{I}{\Delta \phi} = \frac{1}{2\pi} \frac{|t(\mu)|^2}{|r(\mu)|^2},
\end{equation}
for the so called 4-point conductance, a result already mentioned also within
current constraining schemes~\cite{Ng92,Bokes03} and in agreement with
the LRS and therefore it is also compatible with the linear regime
of the Landauer-B\"{u}ttiker formalism.

To summarize this section, incorporating a decoherence caused by many-body
interactions, the invariant current approach does lead to a non-zero
induced drop and the resulting conductance is identical to that obtained
within NEGF formalism as well as the LRS. We should stress
that the way we have incorporated the decoherence contains assumptions
regarding the coupling of the states to the environment and therefore
can not be regarded as a truly {\it ab initio} approach.
The second important outcome is that formulating the maximum-entropy
current-constrained scheme that starts at the non-interacting or mean-field
self-consistent field level will not give physically meaningful results.

\section{The uniform-current theory}
\label{sec-3}

In the uniform-current theory~\cite{Kosov01,Kosov02,Kosov03} one takes
the constraint to make the current uniform throughout the system
\begin{equation}
        I = I(x) = \langle \hat{j}(x) \rangle.
         \label{eq-UC-const}
\end{equation}
This is achieved by introducing a continuous Lagrange multiplier
$\mathcal{A}(x)$ and the corresponding DM takes the form
\begin{equation}
         \hat{\rho}^{UC} = \exp\left\{ \Omega^{UC} - \beta\left( \hat{H} -
         \mu \hat{N} - \int dx \mathcal{A}(x) \hat{I}(x) \right) \right\}
\end{equation}
The function $\mathcal{A}(x)$ has to be found such that the constraint
(\ref{eq-UC-const}) holds. The essential difference from the
invariant-current scheme is the fact that this ansatz results in a density
matrix
which evidently does not commute with the physical Hamiltonian
\begin{equation}
         \left[ \hat{\rho}^{UC} , \hat{H} \right] \neq 0.
\end{equation}
As a consequence, even though the current is uniform at some instant,
(and therefore from the continuity equation the density is momentarily
stationary), it will in general change at later times
together with many other averages computed using the $\hat{\rho}^{UC}$.

Apart from this objection, one can show that the scheme is equivalent
to a equilibrium-like calculation with some effective potential, using
a gauge transformation.  Similarly to the invariant current theory,
it emerges that the induced drop in electrostatic potential is zero. We will
demonstrate
this at the mean-field level of approximation only, even though it presents
no complication in this case to prove it for fully interacting electrons
as well. We firstly reformulate the many-particle problem of the full density
matrix $\hat{\rho}^{UC}$ into that of one-body density matrix, in a
way completely analogous to the previous section and the formalism given
in the Appendix \ref{sec-app-1}. Eventually we will be concerned with
the mean-field single-particle Hamiltonian of the form
\begin{equation}
         \hat{h} =
         -\frac{1}{2} \partial_x^2 + V(x) + \int d^3 x' \mathcal{A}(x')
\hat{I}(x')
         = \frac{1}{2} \left( -i \partial_x + \mathcal{A}(x) \right)^2 + V(x)
           - \frac{1}{2} \mathcal{A}^2(x),
\end{equation}
where $\mathcal{A}(r)$ is the same Lagrange multiplier as used in the
many-particle density matrix, the current operator is
\begin{equation}
         \hat{I}_x(x') = -\frac{i}{2} \left( \partial_x \delta(x-x')
                 + \delta(x-x') \partial_x \right),
\end{equation}
where the subscript $x$ reminds that the operator operates on functions
of $x$. In this way the non-equilibrium problem has been transformed into
a complex (so that current flow is possible in an effectively equilibrium-like
system) but Hermitian eigenvalue problem on the whole space
$x\in (-\infty,\infty)$ (so that the eigenstates form a continuum)
\begin{equation}
         \hat{h} \chi_{E,\alpha}(x) = E \chi_{E,\alpha}(x).
\end{equation}
The final one-particle density matrix is given by analogy with the
previous section by the expression
\begin{equation}
         n(x,x') = \sum_{\alpha} \int dE \frac{\chi_{E,\alpha}(x)
         \chi^{*}_{E,\alpha}(x')}{e^{\beta \left( E - \mu \right) } + 1}.
\end{equation}
However, the complex character of the problem can be removed by a simple
gauge transformation
\begin{equation}
         \chi_{E,\alpha}(x) = e^{-i\int^x \mathcal{A}(x') dx'}
\phi_{E,\alpha}(x),
\end{equation}
where the transformed states $\phi_{E,\alpha}$ are eigenstates of a real
Hamiltonian
\begin{equation}
         \hat{h'} =  \label{eq-h-gauged}
          - \frac{1}{2} \partial_x^2 + V(x) - \frac{1}{2} \mathcal{A}^2(x).
\end{equation}
From this follows that the one-body density matrix
\begin{equation}
         n(x,x') = \sum_{\alpha} \int dE e^{-i\int^x \mathcal{A}(\xi) d\xi}
                 \frac{ \phi_{E,\alpha}(x) \phi^{*}_{E,\alpha}(x')
                         }{e^{\beta \left( E - \mu \right) } + 1}
                 e^{i\int^{x'} \mathcal{A}(\xi) d\xi}, \label{eq-UC-Fermi}
\end{equation}
gives non-zero current only through the gauge-factors. The current
can be therefore easily evaluated to give
\begin{equation}
         \langle I(x_0) \rangle
                 = \int dx dx' \delta(x-x') \hat{I}_x(x_0) n(x,x')
                = \mathcal{A}(x_0) n(x_0),
\end{equation}
where $n(x)$ is the electronic density. We can now fulfill the requirement
of uniformity of the current by giving the Lagrange multiplier as
\begin{equation}
         \mathcal{A}(x) = \frac{I}{n(x)}, \label{eq-A-I}
\end{equation}
which makes the similar results of the previous work by Kosov~\cite{Kosov02}
completely general (the same result holds for interacting electrons since the
gauge transform argument does not depend on the interactions). Using the
result (\ref{eq-A-I}) within the effective Hamiltonian (\ref{eq-h-gauged})
gives a simple closed set of equations to be solved.

Finally we turn to the analysis of the induced drop in the electrostatic
potential which, similarly to the case of the invariant current theory,
is identically zero. To show this we note that the self-consistent
potential $V(x)$ used to determine the wavefunctions and therefore via
Eq.~(\ref{eq-UC-Fermi}) the density does not have any finite drop as
it corresponds to a fictitious equilibrium system (the effective Fermi
function in (\ref{eq-UC-Fermi}) depends only on the energy of
the single-particle wavefunction $\phi_{E,\alpha}(x)$.) for which the drop
must be clearly zero. This observation is not affected by the presence
of the last term in the effective Hamiltonian (\ref{eq-h-gauged}) since
in the case of identical electrodes
\begin{equation}
         \mathcal{A}(x\rightarrow\infty) - \mathcal{A}(x\rightarrow-\infty)
         = I \left( 1/n_B - 1/n_B \right) = 0.
\end{equation}
and the contribution so the drop in potential is zero.

\section{Conclusions}
\label{sec-4}
In conclusion, we have shown how the maximum-entropy formalism
can be applied for non-equilibrium steady states within the framework of the
single-particle approximation. We have presented three different
approaches within a common formalism: (1) the invariant-current constraint, (2)
the invariant-current constraint without the off-diagonals and (3) the
uniform-current constraint scheme. For these we have rigorously derived
the one-particle density matrix for an infinite system that cannot be realized
as a limit of a finite system. Subsequently we have obtained
the expression for the electrostatic drop between electrodes within
the linear response regime. We have shown that (1) and (3) give zero
induced drop in electrostatic potential which is not compatible with actual
current-carrying situations in nano-junctions.
In the case (2) we have shown that by removing the off-diagonal elements of
the current operator, the induced drop is nonzero and in fact, the results
in the linear response are identical to the LRS. We can view the three
different schemes analyzed in the two last sections
from a more general perspective. They represent a variational prescription
for the search of the non-equilibrium state with a given average
of total current. The key difference between them is the part
of current operator that is being used for the constraint. The
uniform-current theory takes the whole, unmodified current operator;
the invariant-current scheme (through the stationarity requirement) removes
the off-diagonals between states of different energy; and finally there is
the form of invariant-current theory in which the off-diagonal elements
within the scattering states basis set representation are removed by
decoherence. Interestingly, only
the last gives a electronic density which results in a nonzero
induced drop in the potential.
This observation rises a question whether there is something inherently
wrong with demanding the current to be fixed. The theory with no
off-diagonal elements of the current operator shows that the explanation
for the problems with the current-constraining schemes arise from the fact
that many-particle interactions or a certain source of decoherence
is essential for the density matrix to give a physically meaningful
results.
\begin{acknowledgments}
This work was supported by the EU's 6th Framework Programme
through the QuaTraFo project (MERG-CT-2004-510615), the Slovak grant agency 
VEGA (project No. 1/2020/05), the NATO Security Through Science Programme 
(EAP.RIG.981521), the EU's 6th Framework Programme through 
the NANOQUANTA Network of Excellence (NMP4-CT-2004-500198) and the EPSRC.
\end{acknowledgments}

\appendix \section{Derivation of the one-particle density matrix}
\label{sec-app-1}
To derive the form of the one-particle density matrix we first define
an auxiliary Green's function, in a close analogy with the Matsubara
technique~\cite{Mahan},
\begin{eqnarray}
         G(2,1) &=& - \mathrm{Tr}\left[ \hat{\rho} T\{
\psi(2),\psi^{\dagger}(1)
                 \} \right], \\
         \psi^{\dagger}(1) &=& e^{\hat{K}\tau_{1}} \psi^{\dagger}(x_{1})
                 e^{-\hat{K}\tau_{1}},
         \psi(2) = e^{\hat{K}\tau_{2}} \psi(x_{2}) e^{-\hat{K}\tau_{2}} \\
         \hat{K} &=& \hat{H} - \mu \hat{N} - A \hat{I}^{0}, \label{K-Ham}
\end{eqnarray}
where $\psi^{\dagger}(x)$ and $\psi(x)$ are the field operators
for electrons. The fictitious time dynamics is given by the effective
many-particle Hamiltonian $\hat{K}$, such that one can employ the similarity
between the unitary time evolution operator and the density matrix
$\hat{\rho}$~(\ref{final-DM}).
The Green's function defined in this way satisfy the equation of motion
\begin{equation}
        \left( -\frac{1}{2} \partial_x^{2} + V(x) - \mu - A I^{0}(x)
        + \partial_{\tau} \right)G(x,x') = - \delta(x-x'), \label{G-eq}
\end{equation}
with the fermionic boundary condition
\begin{equation}
         G(\tau=0,\tau') = - G(\tau=\beta,\tau').
\end{equation}
We note that while the many-body effective Hamiltonian $\hat{K}$ in
(\ref{K-Ham}) is infinite and therefore just a formal expression,
the one-body Hamiltonian present in the equation of motion for the Green's
function (\ref{G-eq}) is finite, for it represents energy {\it per} particle.
The scattering states $\phi_{E,\nu}(x)$ (\ref{eq-scatt-r}), (\ref{eq-scatt-l})
diagonalize the Hamiltonian $-\frac{1}{2} \partial^{2}_{x} + V(x)$ and
therefore
leave the elements of $I^{0}$ in a block-diagonal (diagonal with respect
to the energy) form given by~\cite{Bokes03}
\begin{eqnarray}
        \label{current-matrix}
        I^{0}_{\nu,\eta}(E,E')  \times A &=&  \lim_{T\rightarrow \infty}
                 \frac{1}{2T} \int_{-T}^{T} dt
                 \langle \phi_{E,\nu} | \hat{I}(x,t)| \phi_{E,\eta} \rangle
                 \times A \\
                 &=& \frac{d I_{\nu,\eta}}{dE}(x<<0) \times \delta(E-E')
                 \times \tilde{A},  \label{current-matrix-2}
\end{eqnarray}
where we have used the renormalization of the current-related thermodynamic
parameter $A$ into $\tilde{A}= \lim_{T \rightarrow \infty} \pi A/T$.
The matrix $dI_{\nu,\eta}/dE$ is the matrix of current operator at energy
$E$ given by Eq.~(\ref{eq-curr-1}). The equation of motion for $G(\tau)$
in the scattering-states representation now takes the form
\begin{eqnarray}
                \sum_{\nu'} \left[ \label{eq-of-motion-RL}
                (E-\mu) \delta_{\nu,\nu'}
                - \tilde{A} \frac{d I_{\nu,\nu'}}{dE}
                + \partial_{\tau} \right] G_{E,E''}^{\nu',\eta}(\tau,\tau') =
        - \delta(\tau - \tau') \delta(E-E'') \delta_{\nu,\eta},
\end{eqnarray}
where
\begin{eqnarray}
         G_{E,E'}^{\nu,\eta}(\tau,\tau') = \langle  \phi_{E,\nu} |
         G(x,x') | \phi_{E',\eta} \rangle .
\end{eqnarray}
It is now clear that to solve (\ref{eq-of-motion-RL}) we finally need
to diagonalize the current matrix, i.e. employ the unitary transformation
$U_{\alpha,\nu}$ discussed in the section (\ref{sec-2}). The equation
of motion after this transform takes a simple from
\begin{equation}
        \left( \tilde{E}_{\alpha}  - \partial_{\tau} \right)
         G^{\alpha,\beta}_{E,E'}(\tau) =   \label{eq-G-in-eigen}
                 - \delta(\tau-\tau') \delta_{\alpha,\beta} \delta(E-E').
\end{equation}
We will drop the energy $E,E'$ and state $\alpha,\beta$ indices for a moment
since all the following manipulations are diagonal with respect to these.
Regarding the time, the Green's function must obey the fermionic boundary
condition
\begin{equation}
        G(0) = - G(-\beta).
\end{equation}
The general solution of Eq.~(\ref{eq-G-in-eigen}) is clearly
\begin{equation}
        G = \left\{ \begin{array}{ll}
                C e^{-\tilde{E}(\tau+\beta)} & -\beta < \tau < \tau' \\
                -C e^{-\tilde{E}\tau} & \tau' < \tau < 0
        \end{array} \right.
\end{equation}
Integrating the equation of motion over $\int_{\tau'-\delta}^{\tau'+\delta}$
we see that $G(\tau)$ must have a unit step discontinuity at $\tau=\tau'$
so that we can fix the constant $C$
\begin{equation}
        C e^{-\tilde{E}(\tau'+\beta)} - ( -C e^{-\tilde{E}\tau'}) = 1
\end{equation}
and therefore
\begin{equation}
        C = \frac{e^{\tilde{E} \tau'}}{1 + e^{-\tilde{E}\beta}}
\end{equation}
The density matrix is then given by
\begin{equation}
         G(\tau,\tau') =
         \frac{e^{\tilde{E} \tau'}}{1 +
                 e^{-\tilde{E}}} \times
          \left\{ \begin{array}{ll}
         e^{-\tilde{E}(\tau + \beta)} & \beta < \tau < \tau' \\
         - e^{-\tilde{E}\tau} & \tau' < \tau < 0
         \end{array} \right.
\end{equation}
and the density matrix through the prescription is
\begin{equation}
        \hat{n} = G(\tau =0^{-},0) =
         \frac{ e^{-\tilde{E}0^-}}{e^{\beta \tilde{E}} +1}.
\end{equation}
Restoring all the indices we finally have
\begin{eqnarray}
         n(x,x') &=& G(x,\tau=0^{-};x',
                 \tau'=0) =  \label{eff-Fermi-app}
         \sum_{\alpha} \int dE \frac{\chi_{E,\alpha}(x)
         \chi^{*}_{E,\alpha}(x')}{e^{\beta \tilde{E}_{\alpha}(E)} + 1}.
\end{eqnarray}
Which is the result (\ref{eff-Fermi}) given in the section (\ref{sec-2}).

The above given derivation also shows that our approach does not depend
on the choice of the normalization. If we had chosen the scattering states
normalized to a delta-function of $k$ instead of $E$, the current matrix
elements (\ref{eq-curr-1}) would have to be multiplied by $k$.
However, the delta-function of energy in (\ref{eq-curr-1}) is unchanged,
since it comes from the general considerations of stationarity of
the ensemble~\cite{Bokes03}, so that we have $k\delta(E-E') = \delta(k-k')$.
Using this in the equation of motion for $G$ we see, that the final result
is the same, i.e. it is again the matrix elements of current in the
energy-normalized states that appear in the effective dispersion relation.

\section{Time-reversal symmetry and the maximum entropy states}
\label{sec-app-2}

In this Appendix we show that the states $\chi_+(x)$ and $\chi_-(x)$
are related by the time-reversal symmetry
\begin{equation}
         T \chi_+(x) = \chi^*_+(x) = e^{i\phi} \chi_{-}(x). \label{eq-time-rev1}
\end{equation}

Let $\hat{h}^0 = \hat{h} + AI^0$ be the physical one-particle Hamiltonian
in the Eq.~(\ref{eq-h1}). Next, let $\ket{\chi_{+/-}}$ are the single-particle
states that diagonalize both $\hat{h}^0$ and $I^0$. We will use $T$ for
the time reversal operator which is simply a complex conjugation.
We have
\begin{eqnarray}
         \hat{h}^0 \ket{\chi_+} &=& e \ket{\chi_+}  \\
         \hat{h}^0  T \ket{\chi_+} &=& e T \ket{\chi_+},
                 \textrm{ since } [H^0,T]=0  \label{eq-a2-2} \\
         I^0 \ket{\chi_+} &=& + i \ket{\chi_+}  \\
         I^0 T \ket{\chi_+} &=& - i T \ket{\chi_+},
                 \textrm{ since } T I^0 = - I^0 T  \label{eq-a2-4}
\end{eqnarray}
The last property of the current operator is true for representation
of operators and wavefunctions in a real space, where complex conjugation
of the current operator changes it's sign. Identical statements hold
for the left-current-carrying maximum-entropy state $\ket{\chi_-}$ only with a
reversed
sign of the current eigenvalue $i$.

We know that $\ket{\chi_+}$ and $\ket{\chi_-}$ are degenerate eigenstates
with an eigenvalue $e$. From Eq.~(\ref{eq-a2-2}) we see that $T \ket{\chi_+}$
is also a state degenerate with them, and from Eq.~(\ref{eq-a2-4}) that
it has the same current eigenvalue as $\ket{\chi_-}$. Since this exhausts
the possible degeneracy (2-fold in 1D), the only possibility is
that
$$
         T \ket{\chi_+} = e^{i\phi} \ket{\chi_-},
$$
where $\phi$ is an arbitrary phase factor. This therefore shows
that $\ket{\chi_+}$ and $\ket{\chi_-}$ are related by time-reversal
symmetry.

\bibliographystyle{prsty}
\end{document}